\documentstyle[preprint,aps,prl]{revtex}

\newcommand{\be}{\begin{equation}}
\newcommand{\ee}{\end{equation}}
\newcommand{\sst}{\scriptscriptstyle}

\newcommand{\bea}{\begin{eqnarray}}
\newcommand{\eea}{\end{eqnarray}}
\def\r0{r_{\sst 0}}

\begin{document}
\draft
\title{\v{C}erenkov radiation by charged particles in an
external gravitational field}

\author{\bf Anshu Gupta$^\ddagger$, Subhendra Mohanty$^{\dagger} $
and Manoj K. Samal$^{\#} $}

\address{{\it Theory Group, Physical Research Laboratory, \\
Ahmedabad - 380 009, India }}

%\date{July 1995 - Preliminary draft}

\maketitle

\begin{abstract}

{Charged particles in the geodesic trajectory of
an external gravitational field do not emit electromagnetic
radiation. This is expected from the application of the
equivalence principle. We show here that charged particles
propagating in an external gravitational field with non-zero
components of the Ricci tensor can emit radiation by the
\v{C}erenkov process. The external gravitational field acts like
an effective refractive index for light. Since the Ricci tensor
cannot be eliminated by a change of coordinates, there is no
violation of the equivalence principle in this process.}

\end{abstract}
\vskip 1.6cm
\rule{4.3cm}{0.1mm}
\vskip 2mm
\noindent $^{\ddagger}${\sf E-mail}:{\em anshu@prl.ernet.in} \\
$^{\dagger}${\sf E-mail}:{\em mohanty@prl.ernet.in} \\
$^{\#}${\sf E-mail}:{\em mks@prl.ernet.in}
\vskip 5mm
\begin{flushright}
{\sf{PRL-TH-95/12\\ July 1995\\astro-ph/9509022}}
\end{flushright}
\newpage

An accelerated charge emits electromagnetic radiation at
a rate prescribed by the well known Larmour formula \cite{jackson}.
When the acceleration is due to an external magnetic
field then the resulting synchrotron radiation is the most
important mode of energy loss of a charged particle in
astrophysical situations and is observable as pulsar signals \cite{julian}.
Radiation from accretion discs around neutron stars etc. is
caused by collisional acceleration of the charged
particles \cite{salpeter}. However when the acceleration is caused solely by
an external gravitational field then there is no curvature
radiation \cite{boulaware}. This result is consistent with the equivalence
principle according to which, it is always possible to find
local inertial frame where the acceleration due to gravity
disappears. In this inertial frame, the particle has a constant
velocity and will therefore not emit curvature radiation.
A charged particle with a constant velocity however can emit
\v{C}erenkov radiation if there is an external medium such that
the phase velocity of the photons in that medium is less than
the velocity of the particle \cite{ginz}. We show here that the background
gravitational field has an effective refractive index given by
$n_\gamma^2 (k_0)= |g^{0 0}| (1- R^{i}_{\:\: i} /|g^{0 0}| k_0^2 ) $,
where the $i$ summation is over the spatial components of the
Ricci tensor $R^\mu_{\:\:\: \nu}$ and $k_0$ is the photon frequency.
If the quantity $R^{i}_{\:\: i}$ is negative for a certain
metric, then the effective gravitational refractive index of that
metric $n_\gamma (k_0) > 1$ and radiation by the \v{C}erenkov process is
kinematically allowed. According to the equivalence principle it
is possible to find an inertial frame at each point where the
metric is flat. However the Riemann tensor which is a measure of
the relative acceleration between two nearby geodesics \cite{mtw} does
not neccesarily vanish in an inertial frame, and the components
of the Ricci tensor and the curvature scalar are non-zero. Since
the \v{C}erenkov emission is propotional to the Ricci tensor, it
is non-vanishing in the inertial frame of the particle unlike
the usual curvature radiation. An example of a metric that has
such a property is the gravitational field of the (dark + luminous)
matter in our galaxy. The Newtonian potential which gives rise
to the flat rotation curves is given by
$ \phi (r) =-v_c^2 [1- \ln(r/ r_{max})], \:\:\: r<r_{max}
\approx 100 $ kpc; the corresponding gravitational refractive
index in the local inertial frame is $n_\gamma^2 (k_0)= 1 + (v_c^
{\:\:2}/r^2 k_0^{\:\:2})[1 + 4 \ln(r/r_{max})]$, and therefore
the necessary condition for \v{C}erenkov radiation is satisfied.
We show that for a charged fermion with four momentum $p$ and mass
$m$, the rate of energy radiated by the \v{C}erenkov process is
\be
\frac{dE}{dt}= \frac{Q^2 \alpha_{em} (-R^{\hat{i}}_{\:\: \hat{i}})}
{4 \pi}[ \frac{1}{2 m^2} (-R^{\hat{i}}_{\:\: \hat{i}})+ \ln(\frac{2p_
{\hat{0}}}{m})-
\frac{1}{m} (-R^{\hat{i}}_{\:\: \hat{i}})^\frac{1}{2}] .
\ee
where the hats over the indices represent the components in the local
inertial reference frame. For a neutral fermion with mass $m$
and a non-zero magnetic moment $\mu$ the rate of energy radiated
by \v{C}erenkov process is given by
\be
\frac{dE}{dt}= \frac{\mu^2 (-R^{\hat{i}}_{\:\: \hat{i}})^2}
{4 \pi}[\ln(\frac{2p_{\hat{0}}}{m})-\frac{(-R^{\hat{i}}_{\:\:
\hat{i}})^\frac{1}{2}}{m}] .
\ee

Consider the photon emission by \v{C}erenkov process $f(p)
\rightarrow f(p') + \gamma (k)$ in the local inertial frame
of the incoming fermion. The conservation of energy momentum
{\it i.e. }$ p^{\hat {\mu}}= p'^{\hat {\mu}} + k^{\hat {\mu}}$
implies that the \v{C}erenkov radiation angle between ${\bf p}$
and ${\bf k}$ vectors are given by the relation
\be
\cos{\theta}=\frac{g^{i j} p_i k_j}{(g^{i j} k_i k_j)^\frac{1}{2}
(g^{l m} p_l p_m)^\frac{1}{2}} \equiv  \frac{|g^{\hat{0}  \hat{0}}|p_
{\hat{0}} k_{\hat{0}}}{|{\hat{\bf p}}||{\hat{\bf k}}|} (1+ \frac{k^2}
{2 |g^{\hat{0} \hat{0}} |p_{\hat{0}} k_{\hat{0}}});\:\:\: g^{
\hat{\alpha}\hat{\beta}} = diag(-1, 1, 1, 1) .
\ee
A neccessary condition for the photon emission is therefore that
the right hand side of the eqn. (3) is $\le$ 1 or $\ge$ -1. In
order to see if this condition is satisfied we need the dispersion
relations and the  effective refractive indices of fermions $n_f
\equiv |{\bf p}|/ p_0$ and photons $n_\gamma \equiv |{\bf k}|/ k_0$.
In terms of the refractive indices the neccessary condition for the
\v{C}erenkov process can be written as
\be
-1 \le \frac{|g^{\hat{0}\hat{0}}|}{n_\gamma n_f}[1+ (n_\gamma^2-1)
\frac{k_{\hat 0}}{2 |g^{\hat{0}\hat{0}}| p_{\hat 0}}] \le 1.
\ee
For very high energy particles $n_f$ can be arbitrarily close
to one. Then the condition (4) gives the frequency range of
the emitted photon by \v{C}erenkov process.

The propagation of fermions in a curved background is governed
by the Dirac equation
\be
D_{\hat{\alpha}} \Psi = m \Psi ,
\ee
where the covariant derivative $D_{\hat{\alpha}}$ is defined as \cite{utiyama}
\be
D_{\hat{\alpha}} =e_{\hat{\alpha}}^{\:\:\: \mu} (\partial_\mu + \frac{i}{2}
\omega_\mu^{\:\:\: {\hat{\lambda}}{\hat{\beta}}} \sigma_{{\hat{\lambda}}
{\hat{\beta}}}),
\ee
where $e_{\hat{\alpha}}^{\:\:\: \mu}$ are the tetrads connecting
the local inertial frame coordinates $x^{\hat{\alpha}}$ with the coordinate
frame $x_\mu$ and $\sigma_{{\hat{\alpha}}{\hat{\beta}}}= \frac{i}{4}
[\gamma_{\hat{\alpha}}, \gamma_{\hat{\beta}}]$.
The spin connections $\omega_\mu^{\:\:\: {\hat{\alpha}}{\hat{\beta}}}$ can
be expressed in terms of the tetrads as
\be
\omega_\mu^{\:\:\: {\hat{\alpha}}{\hat{\beta}}}= e^{{\hat{\alpha}} \nu}
e^{\:\:\:
{\hat{\beta}}}_{\nu \:\:\:\: ; \mu}\:\:\:.
\ee
 From eqn. (5), we obtain the anticommutator relation
\be
\frac{1}{2} \{\gamma^{\hat{\alpha}} D_{\hat{\alpha}}, \gamma^{\hat{\beta}}
D_{\hat{\beta}}\} \Psi = m^2 \Psi.
\ee
Using eqn. (6) and simplifying the anti commutator bracket in
eqn. (8), we obtain the wave equation for the propagation of the
fermions in a curved background
\be
[g^{\mu \nu} \partial_\mu \partial_\nu - \frac{1}{4}R -m^2] \Psi
=0,
\ee
where we have used the identity \cite{eguchi}
\be
R=R^{{\hat{\alpha}}{\hat{\beta}}}_{\:\:\:\:\:\:\:  {\hat{\alpha}}{\hat{\beta}}}
=e^{\mu}_{\: \: {\hat{\alpha}}} e^{\nu}_{\:\: {\hat{\beta}}}[\partial_{\mu}
\omega_{\nu}^{\:\:\:\:  {\hat{\alpha}}{\hat{\beta}}}-\partial_{\nu}\omega_
{ \mu}^{\:\:\:\: {\hat{\alpha}}{\hat{\beta}}}-\omega_{\mu}^{\:\:\:\:
{\hat{\lambda}}{\hat{\alpha}}}
\omega_{\:\:\:\: \nu {\hat{\lambda}}}^{\hat{\beta}}+\omega_{\nu}^
{\:\:\:\: {\hat{\lambda}}{\hat{\alpha}}}\omega_{\:\:\:\:
\mu {\hat{\lambda}}}^{\hat{\beta}}] ,
\ee
for the curvature scalar in term of the spin connections and tetrads.
The dispersion relation for the fermions is
\be
g^{\mu \nu}p_{\mu}p_{\nu}+\frac{1}{4} R+m^2=0 ,
\ee
and the effective refractive index of the curved background is
given by
\be
n_{f} = \frac{|{\bf p}|}{p_0} = |g^{00}|^{\frac{1}{2}} (1 -
\frac{m^2 +R /4}{|g^{00}|p_0^{\:\:2}})^{\frac{1}{2}} .
\ee

Turning to photons, the curved space lagrangian
\be
L = \sqrt{-g} g^{\mu \alpha} g^{\nu \beta} F_{\mu \nu} F_{\alpha \beta},
\ee
gives the wave equation
\be
g^{\mu \nu} \nabla_{\mu} \nabla_{\nu} A^{\alpha} - R_{\:\: \mu} ^{\alpha}
A^{\mu} = 0,
\ee
after imposing the gauge condition
\be
\nabla_{\mu} A^{\mu} = 0.
\ee
In the eikonal approximation the corresponding wave equation is
given by
\be
g^{\mu \nu} k_{\mu} k_{\nu} A^i + R^i_{\:\: j} A^j = 0,
\ee
and the dispersion relation is obtained from the condition
\be
det \left (\begin{array}{ccc}
k^{\mu}k_{\mu}+R^1_{\:\: 1} & R^1_{\:\: 2} & R^1_{\:\: 3} \\
R^2_{\:\: 1} & k^{\mu}k_{\mu}+R^2_{\:\: 2} & R^2_{\:\: 3} \\
R^3_{\:\: 1} & R^3_{\:\: 2} & k^{\mu}k_{\mu}+R^3_{\:\: 3}
\end{array}\right) = 0 .
\ee
The dispersion relation in the leading order in Ricci tensor for
a spherically symmetric metric is
\be
g^{\mu \nu} k_{\mu} k_{\nu} + R^i_{\:\: i} = 0,
\ee
and the corresponding refractive index is given by
\be
n_{\gamma} = \frac{|{\bf k}|}{k_0} = |g^{00}|^{\frac{1}{2}} (1 -
\frac{R^i_{\:\: i}}{|g^{00}|k_0^{\:2}})^{\frac{1}{2}}.
\ee
In order that $n_\gamma > 1$ we must have $ R^i_{\:\: i} < 1$. Substituting
eqn. (12) and eqn. (19) in eqn. (4), we find that the \v{C}erenkov
angle in a local inertial frame is given by
\be
\cos{\theta}= [1+ \frac{R^{\hat i}_{\:\: \hat i}}{2 k_{\hat 0}^2}
(1 - \frac{k_{\hat 0}}{p_{\hat 0}}) + \frac{1}{2 p_{\hat 0}^2}
(\frac{1}{4} R + m^2)].
\ee

The rate of energy radiated as photons in the \v{C}erenkov radiation
process $f(p) \rightarrow f(p') + \gamma(k)$ in the local inertial
frame of the incoming fermion is given by

\begin{eqnarray}
\frac{dE}{dt}& =& \frac{1}{2p_{\hat{0}}} \int \frac{d^3 \hat{k}}{2
k_{\hat{0}}}
\frac{d^3 \hat{p'}}{2 p'_{\hat{0}}} \:\:\frac{1}{(2 \pi)^2}
\:\:\delta^{(4)}(\hat{p}-\hat{p'}-\hat{k})\:\: |{\cal M}|^2
k_{\hat{0}} \nonumber\\
             &=&  \frac{1}{2p_{\hat{0}}} \int \frac{d^3 \hat{k}}{2 k_
{\hat{0}}} \frac{d^4 \hat{p'}}{(2 \pi)^2} \:\: \theta(p'_{\hat{0}})
\:\: \delta(\hat{p'}^2-m^2)\:\:
\delta^{(4)}(\hat{p}-\hat{p'}-\hat{k}) \:\:|{\cal M}|^2 k_{\hat{0}} ,
\end{eqnarray}
where for the fermion with charge $Q$,
\be
|{\cal M}|^2=4 Q^2 \alpha_{em}[\frac{
n_\gamma^{2}-2}{n_\gamma^{2}} (p'^{\hat{\alpha}}p_
{\hat{\alpha}} - m^2) + \frac{2 p'^{\hat{\alpha}}k_{
\hat{\alpha}}}{n_\gamma^{\:\: 2}k_{\hat{0}}^{\:\: 2}}
(p_{\hat{0}}k_{\hat{0}} -p^{\hat{\alpha}}k_{\hat{\alpha}}) + \frac{2
p^{\hat{\alpha}}k_{\hat{\alpha}}}{n_\gamma^{\:\: 2} k_{\hat{0}}}
p'_{\:{\hat{0}}} +\frac{n_\gamma^{\:\: 2} -1}{n_\gamma^{\:\: 2}} p'_{\:
\hat{0}}p_{\hat{0}}].
\ee

Writing the mass-shell condition for the outgoing fermion as
\be
\delta(p'^{\hat{\alpha}}p'_{\hat{\alpha}}-m^2)= \frac{1}{2 |{\hat{\bf p}}|
|{\hat{\bf k}}|} \:\:
\delta(\frac{k^2-2 p^{\hat{0}} k_{\hat{0}}}{2 |{\hat{\bf p}}||{\hat{\bf k}
}| }- \cos{\theta}),
\ee
we have
\be
\frac{dE}{dt}=\frac{Q^2 \alpha_{em}}{4 \pi n_\gamma^{2}} \int
\frac{{\hat{\bf k^2}} d{\hat{\bf k}}}{p_{\hat{0}} k_{\hat{0}}
|{\hat{\bf p}}||{\hat{\bf k}}|} \int^1_{-1}d{\cos{\theta}}
\:\:\delta(\frac{k^2-2 p^{\hat{0}} k_{\hat{0}}}
{2 |{\hat{\bf p}}||{\hat{\bf k}}| }- \cos{\theta}) [p_{\hat{0}}
(p_{\hat{0}}-k_{\hat{0}}) \frac{(n_\gamma^{\: 2} - 1)}{n_\gamma^{\:\: 2}}
+ \frac{k^2}{2 n_\gamma^{\:\: 2}}] k_{\hat 0}.
\ee
 From eqn. (21) it is clear that the integral is non-zero when the
criterion (20) is obeyed. The condition (20) therefore gives the
range of the emitted photon frequency in terms of the incoming
fermion energy and the refractive indices of the fermion and the
photon in the external gravitational field. Performing the
integral over the $\delta$- function in (21) and substituting
for $k$ in terms of the refractive index $n_\gamma$, we have the
expression for the energy radiated by a charged particle in a
background gravitational field by the \v{C}erenkov process given
by
\be
\frac{dE}{dt}= \frac{Q^2 \alpha_{em}}{4 \pi p_{\hat{0}}^{\:\: 2}}\:\:
\int_{\omega_1}^{\omega_2} dk_{\hat{0}} (n_\gamma^{\:\: 2} - 1)
[\frac{k_{\hat{0}}^{\:\: 2}}{2} + p_{\hat{0}}
(p_{\hat{0}} - k_{\hat{0}})] k_{\hat{0}},
\ee
where the range of $k_{\hat{0}}$ is where the criterion (20) is
satisfied. The range of allowed photon frequency in local
inertial frame $(\omega_1, \omega_2)$ is obtained by using (12)
and (19) in (4) as
\be
(\frac{-R^{\hat{i}}_{\:\: \hat{i}}}{4})^\frac{1}{2} \le \: k_{\hat{0}}
 \:\le \:(-R^{\hat{i}}_ {\:\: \hat{i}})^\frac{1}{2} (p_{\hat{0}} / m) .
\ee

 From the expression for the refractive index (19) it is clear that
the radiation rate given by (25) is positive for the background
metrics for which $R_i^{\:\: i} < 0$. The rate of energy loss in
this background through \v{C}erenkov radiation is obtained by
substituting (19) in (25) and using (26) to yield
\be
\frac{dE}{dt}= \frac{Q^2 \alpha_{em} (-R^{\hat{i}}_{\:\: \hat{i}})}
{4 \pi}[ \frac{1}{2 m^2} (-R^{\hat{i}}_{\:\: \hat{i}})+ \ln(\frac{2p_
{\hat{0}}}{m})- \frac{1}{m} (-R^{\hat{i}}_{\:\: \hat{i}})^\frac{1}{2}] .
\ee

Neutral fermions with non-zero magnetic dipole moment like
neutrons and possibly neutrinos can also emit \v{C}erenkov
radiation. The amplitude for the process $f(p) \rightarrow
f(p') + \gamma (k)$ through the magnetic dipole vertex is
given by
\be
{\cal M}=\frac{\mu}{n_\gamma} \sigma^{\mu \nu} \epsilon_\mu k_\nu \:,
\ee
where $\mu$ is the magnetic dipole moment of $f$. Substitution of
$|{\cal M}|^2$ in (21) followed by the integration over the
relevant variables yields \cite{grimus}
\be
\frac{dE}{dt}= \frac{\mu_\nu^2}{16 \pi p_{\hat{0}}^2}
\frac{(n_\gamma^2-1)^2}{n_\gamma^2} \int_{\omega_1}^{\omega_2}
k_{\hat{0}} dk_{\hat{0}} [4 p_{\hat{0}} ^2 k_{\hat{0}} ^2 -4p_{\hat{0}}
k_{\hat{0}} ^3- (n_\gamma^2-1)k_{\hat{0}} ^4].
\ee
Then we found out the rate of energy radiated in this case to be
\be
\frac{dE}{dt}= \frac{\mu^2 (-R^{\hat{i}}_{\:\: \hat{i}})}
{4 \pi}[\ln(\frac{2p_{\hat{0}}}{m})- \frac{(-R^{\hat{i}}_{\:\:
\hat{i}})^\frac{1}{2}}{m}] .
\ee

An example of a metric where the Ricci tensor $R^i_{\:\: i}$
are negative is the gravitational field of the
dark matter distribution in our galaxy. The galactic metric is
described by the components $g_{rr}= (1-2\phi)$, $g_{tt}=
(1+2\phi)$ where $\phi$ is the Newtonian potential
that gives rise to the flat rotation curve and is given by
\begin{eqnarray}
\phi(r) &=&-v_c^2 [1-\ln (\frac{r}{r_{max}})], \:\:\:\: r<r_{max}\nonumber\\
 &=& -\frac{G M_g}{r},
\:\:\:\:\:\:\:\:\:\:\:\:\:\:\:\:\:\:\:\:\:\:\:\:\: r>r_{max},
\end{eqnarray}
where $v_c \approx 220$ km/sec is the rotational velocity of the
galactic halo, $r_{max} \approx 100$ kpc is the extent of the
dark matter halo. The components of the Ricci tensor
corresponding to the Newtonian potentials (28) are
\begin{eqnarray}
R^t_{\:\: t} &= & R^r_{\:\: r} = - \frac{v_c^2}{r^2},\nonumber\\
R^{\theta}_{\:\:\: \theta}& =& R^{\phi}_{\:\:\: \phi} =  - \frac{ 2
v_c^2}{r^2}\:\; \ln(\frac{r}{r_{max}}),
\end{eqnarray}
for $r<r_{max}$ and $R^\mu_{\:\: \nu} =0$ for $r>r_{max}$. The
refractive index of light in its inertial frame in such a gravitational
field is given by
\be
n_\gamma^{\:\: 2}= 1 + \frac{v^2_c}{r^2 k_0^2}[1 + 4 \ln(r/r_{max})],
\ee
which is greater than $1$ as long as $\ln(r/r_{max}) > -0.25 $.
Then the rate of \v{C}erenkov radiation is given by substituting
(32) in (27) to yield
\be
\frac{dE}{dt} =  \frac{Q^2 \alpha_{em}}{4 \pi} \frac{v^2_c}{r^2}
[1 + 4 \ln(r/r_{max})] \ln{\frac{2 p_{\hat 0}}{m}}.
\ee

The total energy loss for extra galactic cosmic rays due to the
\v{C}erenkov process by the time they arrive on the earth at
distance $r$ is given by
\be
\Delta E (r)=\int_r^{r_{max}} \frac{dE}{dt} dr =  \frac{Q^2
\alpha_{em} v_c^2}{4 \pi} \ln(\frac{2 p_{\hat
0}}{m})(\frac{1}{r}- \frac{1}{r_{max}}).
\ee
For cosmic ray protons of energy $10^{19}$ eV, this turns out to be
$10^{-23}$ eV. The energy loss is expected to be large in the
accretion discs of compact sources
and may be of observational significance.

\end{document}